\documentclass[aps,twocolumn,superscriptaddress,footinbib]{revtex4}

 \usepackage{graphics}
 \usepackage{graphicx}
 \usepackage{epsfig}
 
 \usepackage{hyperref}

\usepackage{amssymb}
 \usepackage{amsthm}
 
\usepackage{amsmath,mathtools}
\usepackage[framemethod=TikZ]{mdframed}
\usepackage{rotating}

\newcommand{\vect}[1]{\boldsymbol{#1}}

\begin{document}
\title{The Huygens Principle of Angle-Resolved Photoemission}

\author{Simon Moser}\affiliation{Physikalisches Institut and W\"{u}rzburg-Dresden Cluster of Excellence ct.qmat, Universit\"{a}t W\"{u}rzburg, 97074 W\"{u}rzburg, Germany}
\email[E-mail address: ]{simon.moser@physik.uni-wuerzburg.de}

\begin{abstract}
Angle-resolved photoemission spectroscopy (ARPES) measures the interference of dipole allowed Coulomb wavelets from the individual orbital emitters that contribute to an electronic band. If Coulomb scattering of the outgoing electron is neglected, this Huygens view of ARPES simplifies to a Fraunhofer diffraction experiment, and the relevant cross-sections to orbital Fouriertransforms. This plane wave approximation (PWA) is surprisingly descriptive of photoelectron distributions, but fails to reproduce kinetic energy dependent final state effects like dichroism. 
Yet, Huygens principle of ARPES can be easily adapted to allow for distortion and phase shift of the outgoing Coulomb wave. This retains the strong physical intuition and low computational cost of the PWA, but naturally captures momentum dependent interference effects in systems that so far required treatment at the \textit{ab initio} level, such as linear dichroism in Rashba systems BiAg$_2$ and AgTe.
\end{abstract}

\maketitle 

\textit{Introduction.}---In 1678,  Dutch physicist Christiaan Huygens proposed his famous principle of wave mechanics, stating that \textit{every point on a wavefront is itself the source of a spherical wavelet}. Combined with Augustin-Jean 
Fresnel's 1818ies insight that \textit{these secondary wavelets all mutually interfere to form the actual wavefront}, this intuitive picture provides an appropriate explanation of (near and far field) wave-propagation, reflection and refraction, and most importantly: diffraction \cite{Miller1991}. 

For angle-resolved photoemission spectroscopy (ARPES), a well established technique to map electronic structure of adsorbed molecules and ordered solid state, such an intuitive interpretation in terms of simple wave mechanics remains elusive. This is remarkable, as fundamentally, the electronic structure contrast produced by ARPES relies on the coherent interference of photoelectron wavelets emitted from individual orbital emitters that are phase-locked through their atomic arrangement, i.e., the structure of a particular molecule, or the lattice properties of an ordered solid. 

In 2009, this very insight along with the availability of efficient photoelectron detectors, pioneered a novel imaging technique of real space molecular orbitals based on the ARPES response of adsorbed organic molecules \cite{Puschnig2009,Luftner2014}. This \textit{orbital tomography} technique assumes that the photoelectrons transition into plane waves that freely propagate to the detector, and that the ARPES intensity distribution is determined by the real space orbital's Fourier transform (see Ref.~\cite{MoserME2016} and references therein). ARPES hence intuitively maps onto a Fraunhofer diffraction experiment and the quest for a \textit{Huygens principle of ARPES} thus seems to be complete. 

The tempting use of plane wave final states is flawed, however, as it neglects scattering of the outgoing photoelectron in the Coulomb potential of the ion it leaves behind \cite{Bradshaw2015}, and thus inherently fails to describe photon energy dependent final state interference such as dichroism. In fact, it produces an ubiquitous $\vect{\epsilon}\cdot\vect{k_f}$ polarization term that genuinely suppresses outgoing photoelectron momenta $\vect{k_f}$ that move perpendicular to the polarization vector $\vect{\epsilon}$; a model artifact that is rarely observed with this stringency in experiments.
\\\\
\textit{Model.}---Yet, such discrepancies can be overcome without loss of intuition or computational ease, taking into account the appropriate scattering state of the outgoing photoelectron, i.e., a partial wave expansion

\begin{eqnarray}\label{eq: partial wave expansion}
\chi_{\eta}(\vect{r}) &=& 4 \pi \sum_{l=0}^{\infty} \sum_{m=-l}^{l} i^l~e^{i \sigma_l}R_{\eta l}(r) Y_l^m (\vect{\Omega}_{k_f})Y_l^{m*} (\vect{\Omega}_r)\nonumber
\end{eqnarray}

\noindent in terms of Coulomb wavelets built from spherical harmonics $Y_l^m$, radial wave functions $R_{\eta l}$ and Coulomb phase shifts $\sigma_l$ \cite{Messiah1961,Cooper1962}. The Coulomb distortion of $\chi_{\eta}$ with respect to the free electron is described by the Sommerfeld parameter $\eta=Z/a_0 k_f$, which in the limit of small ion charge $Z$ and large photoelectron momenta $k_f\equiv|\vect{k}_f|$ yields $\eta\rightarrow 0$, $R_{\eta l}(r)\rightarrow j_l(k_f r)$ and $\sigma_l\rightarrow 0$, and thus naturally retrieves the plane wave expansion $\chi_{\eta}(\vect{r}) \rightarrow e^{i \vect{k}_f\cdot\vect{r}}$ \cite{Messiah1961,Abramowitz1988}. 

Computing dipole transitions from a hydrogen like atomic orbital $\Phi_{nlm}$ into scattering states $\chi_{\eta}$, we find

\begin{eqnarray}
\vect{M}^{\eta}_{nlm}&\propto &\langle \chi_{\eta}| \vect{\nabla}|\Phi_{nlm}\rangle\nonumber\\
&=&\underbrace{\widetilde{f}(k_f)~  \vect{Y}_{l,l+1,m}(\vect{\Omega}_{k_f})}_{\text{dipole transition }l\rightarrow l+1} + \underbrace{\widetilde{g}(k_f)~  \vect{Y}_{l,l-1,m}(\vect{\Omega}_{k_f})}_{\text{dipole transition }l\rightarrow l-1}  ~,\nonumber
\end{eqnarray}

\noindent where we introduced the complex-valued radial cross-sections $\widetilde{f}(k_f)$ and $\widetilde{g}(k_f)$, whose atomic limit is given more explicitly in the Suppl. Info.

Based on this expression, we can now formulate the Huygen's principle of ARPES: \textit{Every atomic orbital participating in the photoemission process is the source of two dipole allowed Coulomb wavelets, and the Coulomb wavelets emanating from all these orbital emitters mutually interfere}. More explicitly, the vector spherical harmonics $\vect{Y}_{l,l\pm1,m}$ describe the orbital symmetry of the two dipole allowed emission channels $l\rightarrow l\pm1$ that are reached by a given polarization vector $\vect{\epsilon}$ \cite{Arfken2012}, whose interference is determined by their $k_f$-dependent radial cross-section ratio $|\widetilde{f}|/|\widetilde{g}|$ and relative phase $\Delta\sigma=\arg(\widetilde{f}/\widetilde{g})$.

While in the PWA, both $|\widetilde{f}|/|\widetilde{g}|=\text{const}$ and $\Delta\sigma=0$ are independent of $k_f$ (see Suppl. Info), it is precisely their Coulomb induced $k_f$-dependence that produces kinetic energy dependent final state interferences -- and thus bears the potential to describe photon energy dependent dichroism as we will see in more detail later. 

We note in passing, that this $k_f$-dependence of $|\widetilde{f}|/|\widetilde{g}|$, i.e., the fact that individual photoemission channels can be suppressed or enhanced by an appropriate choice of photon energy, is a direct consequence of the dipole operator's velocity form $\mathcal{H}_{\text{int}}\propto\vect{\epsilon}\cdot\vect{\nabla}$ that we use here, but is not reproduced by its length form $\mathcal{H}_{\text{int}}\propto\vect{\epsilon}\cdot\vect{r}$ \cite{Day2019} (see Suppl. Info). In particular, while the length form is unbounded and not well defined for extended (infinite) systems \cite{Pendry1976,Drake2006,Lebech2012}, the velocity form is translation invariant and thus directly applicable to Bloch states, whose Wannier representation in turn can be expanded in terms of an atomic orbital basis
\begin{equation}
\Psi_{\vect{k}}(\vect{r})=\sum_{\vect{R}} e^{i \vect{k}\cdot\vect{R}}\sum_{nlm}c^{\vect{k}}_{nlm}\Phi_{nlm}(\vect{r})~,\nonumber
\end{equation}
\noindent and the first sum runs over all lattice sites $\vect{R}$ participating in the photoemission process. In the independent center approximation, i.e., ignoring scattering of outgoing electrons in the Coulomb potential of adjacent atoms, the ARPES intensity now compactly reads
\begin{equation}\label{eq: matrix element with phase}
I  \propto  |\langle \Psi_{\vect{k}_f}| \vect{\epsilon}\cdot\vect{\nabla} |\Psi_{\vect{k}}\rangle |^2=\delta(\vect{k}-\vect{k}_f)~| \vect{\epsilon}\cdot \vect{\mathcal{M}_{\vect{k}}}\cdot \vect{c}_{\vect{k}}|^2~,
\end{equation} 

\noindent with the $N \times 3$ dimensional dipole transition matrix $\vect{\mathcal{M}_{\vect{k}}}$ coupling the $N$-dimensional initial state vector $\vect{c}_{\vect{k}}=(c^{\vect{k}}_{n_1l_1m_1},...,c^{\vect{k}}_{n_Nl_Nm_N})^\top$ 
to the three dimensional polarization vector $\vect{\epsilon}$, and $\delta(\vect{k}-\vect{k}_f)$ representing momentum conservation (we thus set $\vect{k}_f \equiv \vect{k}$ from now). 
\\\\
\textit{Dichroism.}---Analyzing matrix equation \ref{eq: matrix element with phase}, we immediately note that exploiting full polarization control in an experiment can yield a maximum of six linear independent equations to retrieve at most three complex eigenvector components of $\vect{c}_{\vect{k}}$ from ARPES intensity measurements. Taking, e.g., a linear combination $|\Psi\rangle=c_x |p_x\rangle+c_y |p_y\rangle+c_z |p_z\rangle$ of $p$-orbitals, we find  
$
\vect{\mathcal{M}}_{\vect{k}}\propto   
\widetilde{f}(k)~\vect{\mathcal{M}}_{p\rightarrow d}(\vect{\Omega}_{k}) 
+ 
\widetilde{g}(k)~\vect{\mathcal{M}}_{p\rightarrow s}(\vect{\Omega}_{k})
$, 

\noindent where the $3 \times 3$ matrices $\vect{\mathcal{M}}_{p\rightarrow d}$ and $\vect{\mathcal{M}}_{p\rightarrow s}$ describe dipole transitions from the initial state $\vect{c}=(c_x,c_y,c_z)^\top$ into $d$ and $s$ channels at a given polarization $\vect{\epsilon}$. Developing these matrices in terms of small photoelectron emission angles $\theta_{k}$ \cite{Arfken2012}, we find (in Cartesian coordinates)

\scalebox{0.67}{\parbox{\linewidth}{
\begin{eqnarray}\label{eq. p-coupling}
\underset{p\rightarrow d}{\vect{\mathcal{M}}(\vect{\Omega}_{k})}&=&
\frac{1}{2 \sqrt{2 \pi }}
\left(
\begin{array}{ccc}
 1 & 0 & 0 \\
 0 & 1 & 0 \\
 0 & 0 &-2 \\
\end{array}
\right)+
\frac{3 }{2 \sqrt{2\pi }} 
\left(
\begin{array}{ccc}
 0 & 0& -\cos\phi_{k} \\
 0 & 0 & \sin \phi_{k} \\
\cos \phi_{k}  & \sin \phi_{k}  & 0 \\
\end{array}
\right)\theta_{k} +
\mathcal{O}(\theta_{k}^2)~;\nonumber\\
\underset{p\rightarrow s}{\vect{\mathcal{M}}(\vect{\Omega}_{k})}&=&\frac{1}{2 \sqrt{\pi }}\left(
\begin{array}{ccc}
 1 & 0 & 0 \\
 0 & 1& 0 \\
 0 & 0 & 1 \\
\end{array}
\right)~.
\end{eqnarray}
}}

\begin{figure*}
\begin{center}
\includegraphics[width=\textwidth]{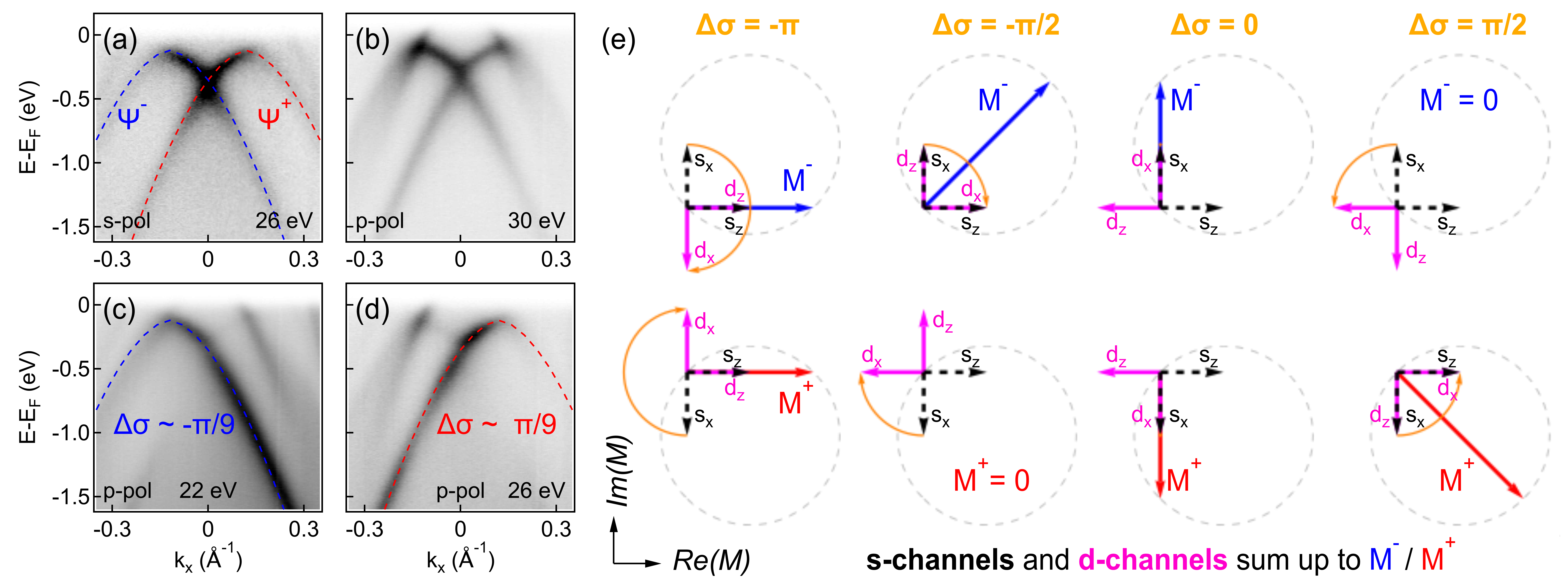}
\caption{(a-d) ARPES data measured along the $k_xk_z$ mirror plane of BiAg$_2$/Ag(111), reproduced from Ref.~\onlinecite{Bentmann2017}. (a) Measurement with $s$-polarized light and $h\nu=26$~eV. Both spin-orbit split surface states $\Psi^{\pm}$ appear with equal intensities $I_s^+\sim I_s^-$ irrespective of photon energy $h\nu$ (see Ref.~\onlinecite{Bentmann2017} for more data). In contrast, ARPES taken with $p$-polarized light exhibits a $h\nu$-dependent intensity swap between $\Psi^{\pm}$: (b) $I_p^+\sim I_p^-$ at $h\nu=30$~eV; (c) $ I_p^+/I_p^-\sim 0$ at $h\nu=22$~eV; (d) $  I_p^-/I_p^+\sim 0$ at $h\nu=26$~eV. (e) Illustration of phase dependent photoemission $s$- ($s_{x/z}$, black dashed) and $d$-channel ($d_{x/z}$, magenta) interference, individually shown for $\Psi^-$ (blue, top) and $\Psi^+$ (red, bottom) in the complex plane. Channels are represented by unit vectors for clarity. For $\Delta\sigma=\pm \pi$ and $0$, the absolute phase between channels $s=s_x+s_z$ and $d=d_x+d_z$ is $\pi/2$ for bands $\Psi^\pm$, and their intensities $I^+=|M^+|^2=|M^-|^2=I^-$are  equal. For $\Delta\sigma=\pm\pi/2$, however, $s$-and $p$-channels of $\Psi^{\pm}$ are in phase and interfere constructively, while they are in antiphase and interfere destructively for $\Psi^{\mp}$.}
\label{fig_BiAg2}
\end{center}
\end{figure*}

\begin{figure}
\begin{center}
\includegraphics[width=\columnwidth]{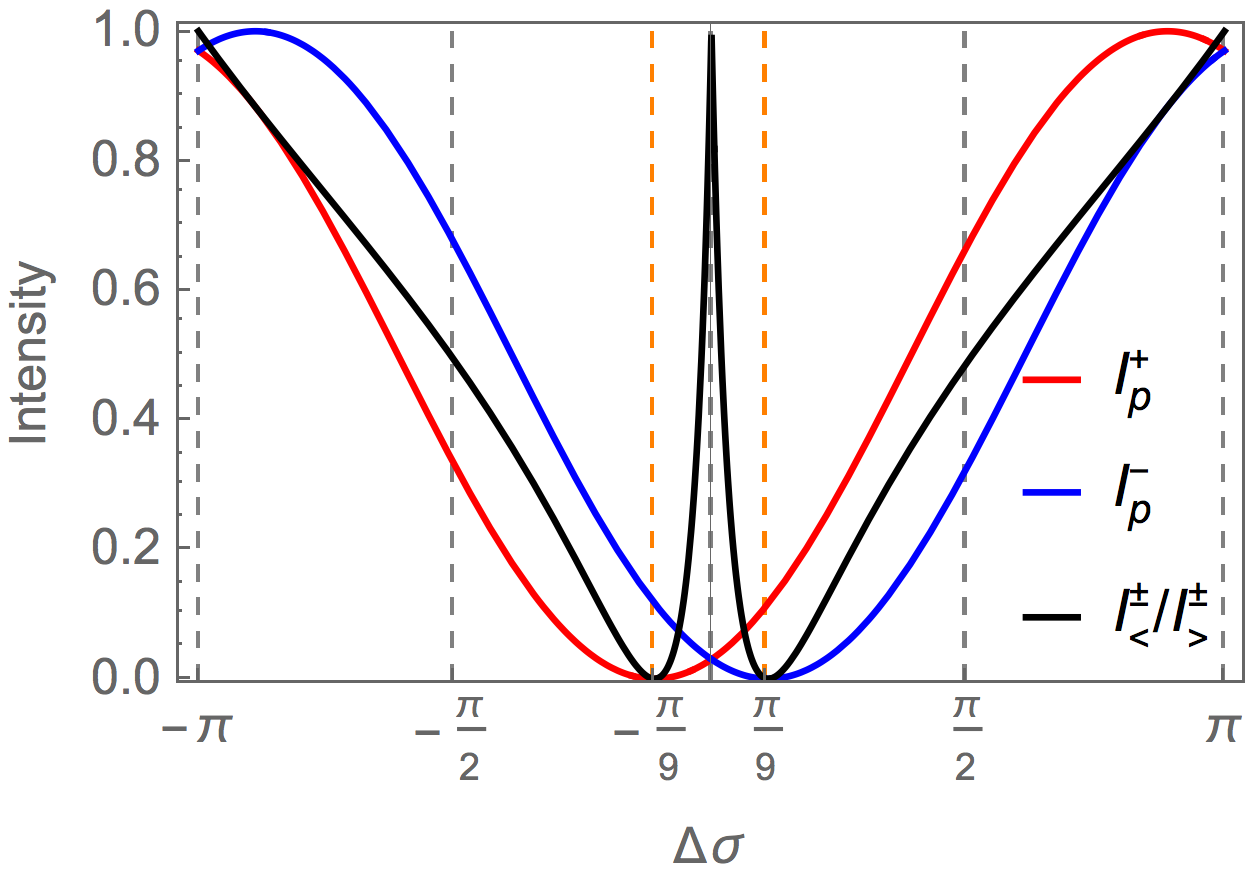}
\caption{Phase dependent ARPES intensity of bands $\Psi^\pm$ in AgBi$_2$ calculated for the explicit experimental geometry $\tan\alpha=\epsilon_z/\epsilon_x=3$ used in Fig.~\ref{fig_BiAg2}~(b-d) \cite{Bentmann2017}. As detailed out in Suppl. Info, we find a unique value set $|\widetilde{f}|/|\widetilde{g}|=0.72$ and $\Delta\sigma \sim \mp\pi/9$ that reproduces the complete intensity suppression of $\Psi^\pm$ in Figs.~\ref{fig_BiAg2} (c) and (d).
 }
\label{fig_BiAg2_2}
\end{center}
\end{figure}

Clearly, the $d$-channel mixes in and out-of plane orbital and polarization components and only diagonalizes right at normal emission $\theta_{k}= 0$, where the contributions from $p_z$ are twice as large and in antiphase to the contributions from $p_x/p_y$. (As we will see later, this has important consequences for bands carrying orbital angular momentum OAM.) In contrast, the $s$-channel is isotropic and diagonal for any emission angle $\vect{\Omega}_k=(\theta_k,\phi_k)$, and thus provides a one-to-one mapping of eigenvector- onto light polarization components. This implies that ARPES at photoelectron momenta where the $d$-channel is suppressed, i.e., $\widetilde{f}(k)=0$, is a direct probe of the band's orbital character. In particular, linear polarizations $\vect{\epsilon}_{x}$ and $\vect{\epsilon}_{y}$ then directly probe the eigenvector amplitudes $|c_x| \propto \sqrt{I_x}$ and $|c_y| \propto \sqrt{I_y}$ (the $c_z$ component is fixed by the normalization condition $c_x^2+c_y^2+c_z^2=1$), while linear (LD) and circular dichroism (CD) in the $xy$-plane probe their mutual interference
\begin{eqnarray}
I_{\text{LD}}&\propto&\frac{I_{x-y} - I_{x+y}   }{ I_{x-y} + I_{x+y}} =\frac{2 \Re (c_x^* c_y)  }{| c_x| ^2+| c_y| ^2}~;\nonumber\\
I_{\text{CD}}&\propto&\frac{I_{x-iy} - I_{x+iy}   }{ I_{x-iy} + I_{x+iy}}=\frac{2 \Im (c_x^* c_y)  }{| c_x| ^2+| c_y| ^2}=\frac{1}{\hbar}\frac{\langle L_z\rangle }{| c_x| ^2+| c_y| ^2}~,\nonumber
\end{eqnarray}

\noindent and retrieve OAM component $\langle L_z\rangle=\langle\Psi|L_z|\Psi\rangle$ and phase relation $ \arg (c_x/ c_y) =\arctan (I_{\text{CD}}/I_{\text{LD}})$.
\\
In analogy, additional LD and CD experiments within the orthogonal $xz$ and $yz$ planes will further provide access to the angular momenta $\langle L_y\rangle$ and $\langle L_x\rangle$ as well as phase relations $\arg (c_{x}/ c_{z})$ and $\arg( c_{y}/ c_{z})$, respectively, in principle allowing for a full reconstruction of the eigenstate vector $\vect{c}$ from ARPES intensity measurements \cite{Schuler2019,Schuler2021}. Note, however, that this generally requires the photoemission $d$-channel and consequent final state interferences to be reliably suppressed, i.e., a photon-energy where $\widetilde{f}(k)\sim0$.
\\\\
\textit{Application.}---Let us illustrate this corollary in a well studied model system whose large $Z$ constituents give rise to elevated final state scattering and strong spin orbit coupling (SOC): the surface alloy BiAg$_2$/Ag(111) \cite{Bentmann2017}. Density functional theory (DFT) finds its low energy surface electronic structure to be of primarily Bi $6p$ and Ag $5s$ orbital character, with two Rashba bands $\Psi^\pm$ whose SOC shaped wave-functions along the system's $k_xk_z$ mirror plane are well described by $|\Psi^\pm\rangle = \frac{1}{\sqrt{2}}|p_z,\uparrow\downarrow\rangle \mp \frac{i}{2}|p_x,\uparrow\downarrow\rangle \pm \frac{1}{2} |p_y,\downarrow\uparrow\rangle$, with spinors $|\uparrow\downarrow\rangle$ quantized along the $y$-axis and orbital angular momenta $\langle L_y\rangle^\pm=\mp\hbar/\sqrt{2}$ (Fig. \ref{fig_BiAg2} a) \cite{Mirhosseini2009,Zhang2013}. \\
ARPES experiments with $s$-polarized light and the sample mirror- and ARPES scattering planes coinciding (Fig. \ref{fig_BiAg2} a), display both the $\Psi^+$ and $\Psi^-$ Rashba bands with equal intensity, irrespective of photon energy (See detailed data in Ref.~\cite{Bentmann2017}). In contrast, ARPES experiments with $p$-polarized light find a photon energy dependent swap of intensity between $\Psi^+$ and $\Psi^-$ (Fig. \ref{fig_BiAg2} b-d) \cite{Meier2009,Bentmann2017}.

Based on our model, these observations can be easily understood in terms of photon energy-dependent $s$ and $d$ channel interference: From Eq.~\ref{eq. p-coupling}, ARPES close to normal emission with $s$-polarized light $\vect{\epsilon}_y$ mostly projects out the $p_y$-orbital contributions, leading to an equal intensity distribution $I_s^\pm\propto |\pm \frac{1}{2}(\sqrt{2}\widetilde{g}+\widetilde{f})|^2$ among both bands $\psi^\pm$, modulating synchronously with the $k$-, i.e. kinetic- or photon energy dependent interference of the $s$- and $d$- channels. 

In contrast, the $p$-polarized geometry is receptive to both the $p_x$ and $p_z$ orbital contributions. The intensity distribution is given by $I_p^\pm \propto|\mp \frac{i}{2}(\sqrt{2}\widetilde{g}+\widetilde{f})\cos\alpha+\frac{1}{\sqrt{2}}(\sqrt{2}\widetilde{g}-2\widetilde{f})\sin\alpha|^2$, where $\alpha$ is the angle of light incidence that quantifies the ratio of in- and out of plane polarization $\tan\alpha=\epsilon_z/\epsilon_x$. Interestingly, we now find a disparity $I_p^+-I_p^-\propto |\widetilde{f}||\widetilde{g}|\sin 2\alpha \sin \Delta\sigma$ between bands $\Psi^\pm$ that scales with the Coulomb phase shift $\Delta\sigma = \arg(\widetilde{f}/\widetilde{g})$ between $s$- and $d$-channels, but vanishes for $|\widetilde{f}||\widetilde{g}|=0$ (either channel suppressed) and $\alpha = 0$ or $\pi/2$ ($\vect{\epsilon}_x$ and $\vect{\epsilon}_z$ not mixed). This is a direct consequence of the interference of \textit{both} the $s$ and $d$-channels resulting from \textit{both} the $p_x$ and $p_z$ orbitals. In particular, the pertinent $\pi$-phase shift between the $p_x$ and $p_z$ derived $d$-channels (the minus sign in the $d$-channel entry `-2' of expression~\ref{eq. p-coupling}) reverses their chirality with respect to the $s$-channels. This along with the opposite OAM $ \langle L_y\rangle^\pm=\mp \hbar /\sqrt{2}$ of bands $\Psi^\pm$ (the $\pm \pi/2$ phase between $p_x$ and $p_z$ orbitals) results in a band dependent phase difference between $s$- and $d$-waves that is controlled by the Coulomb phase shift $\Delta\sigma$.

We visualize this effect in Fig.~\ref{fig_BiAg2} (e), where the $d_{x/z}$-channels emitted from $p_x$ and $p_z$ orbitals (for clarity represented by unit vectors in the complex plane) are rotated by $\Delta\sigma$ around their corresponding $s_{x/z}$-channels. For $\Delta\sigma=\pm \pi$ and $0$, the absolute phase between $s=s_x+s_z$ and $d=d_x+d_z$ is $\pi/2$ for both bands $\Psi^\pm$, and their ARPES intensities $I^\pm\propto |M^\pm|^2=|s_x+d_x+s_z+d_z|^2$ are consequently identical. For $\Delta\sigma=\pm\pi/2$, however, $s$- and $d$-channels are in phase and interfere constructively for $\Psi^\pm$, while they are out of phase and interfere destructively for $\Psi^\mp$. This mutual exchange of intensity thus results from the interplay between the opposite chiralities of bands $\Psi^\pm$ (their OAMs; marked by the sign change in $s_x$) in concert with the opposite chiralities of their photoemission $s$- and $d$-channels (marked by the sign change in $d_z$). According to Eq.~\ref{eq. p-coupling}, these arguments also hold for systems carrying OAM along $x$ and light polarized in the $yz$-plane, while the effect vanishes for OAM along $z$ and $xy$-polarized light, where $s$- and $d$-channel chiralities are equal (Suppl. info). 

Returning to AgBi$_2$ and examining $I_p^\pm $ within the experimental geometry $\epsilon_z/\epsilon_x\sim 3$ used in Ref.~\cite{Bentmann2017} in more detail (Suppl. Info), we identify a unique cross-section ratio $|\widetilde{f}|/|\widetilde{g}|\sim 0.72$ that produces the total band suppression $I_p^\pm/I_p^\mp\sim 0$ observed in Fig.~\ref{fig_BiAg2} (c,d) for a phase shift of $\Delta\sigma \sim \pm \pi/9$ (Fig.~\ref{fig_BiAg2_2}). For these particular photon energies, the model thus allows us to extract detailed information on the photoemission final state.

Let us further study the angular dependence of linear dichroism in a system with similar out- and in plane orbital mixing: the 2D honeycomb monolayer AgTe/Ag(111) \cite{Unzelmann2020}. Its occupied low energy electronic structure is described by two distinct bands: band $|\alpha\rangle = \sin\phi_k|p_x\rangle -\cos\phi_k | p_y\rangle$, of tangential orbital character with zero angular momentum; and Rashba split band $|\beta\rangle = \cos\phi_k|p_x\rangle +\sin\phi_k | p_y\rangle + i k_{\|} \delta_{sp}|sp_z\rangle$ \footnote{The splitting is not resolved in the data shown.}, of primarily radial orbital character, but with in- and out of plane orbital mixing $\delta_{sp}\sim 4.15$~\AA~producing a rotating in plane orbital angular momentum $\langle \vect{L}\rangle^\beta=2k_{\|} \delta_{sp}\hbar ~(\sin \phi,\cos\phi,0)^\top$ that is governed -- in contrast to BiAg$_2$ where SOC is decisive \cite{Bentmann2017} -- by inversion symmetry breaking at the surface \cite{Unzelmann2020,UenzelmannPhD}.

\begin{figure}
\begin{center}
\includegraphics[width=\columnwidth]{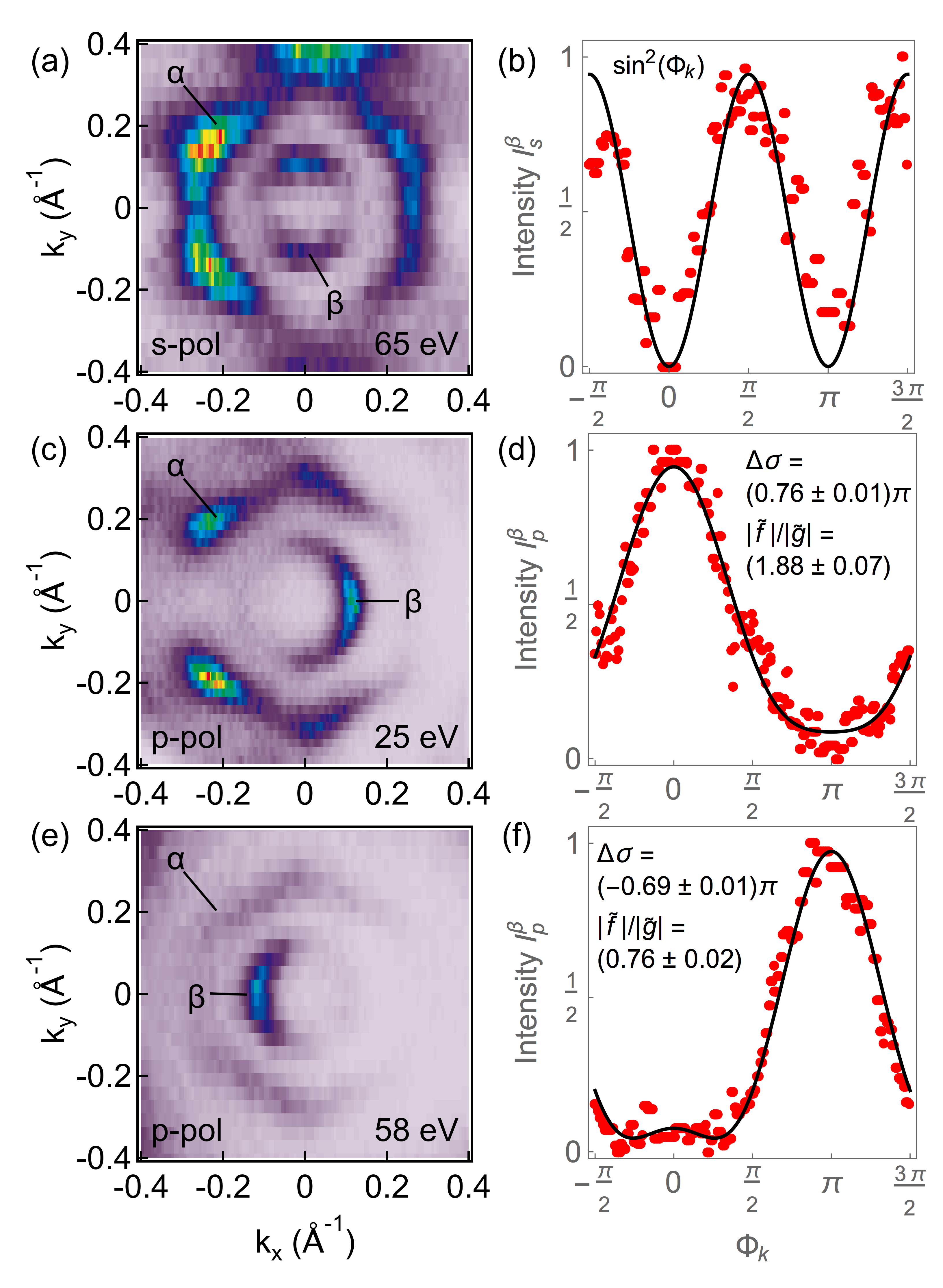}
\caption{(a) AgTe/Ag(111) ARPES constant energy maps at 1.3~eV binding energy, reproduced from Ref.~\onlinecite{Unzelmann2020}. As described in the text, band $\beta$ exhibits a two-fold rotational symmetry when measured with $s$-polarized light at $h\nu=65$~eV (a,b), but shows a distinct half moon signature with $p$-polarized light at $25$~eV (c,d), which is flipped at $58$~eV (e,f). Black curves in (b,d,f) show a best fit to angular intensity distributions extracted from (a,c,e) at $k_\|=0.12$~\AA$^{-1}$, employing the experimental geometry $\tan\alpha=1/\sqrt{2}$ (magic angle light incidence) and the orbital mixing parameter $\delta_{sp}\sim 4.15$~\AA~of Ref.~\onlinecite{Unzelmann2020}}
\label{fig_AgTe}
\end{center}
\end{figure}

Like in our previous discussion, $s$-polarized light close to normal emission projects out the $p_y$ character and delivers $I_s^\alpha\propto \cos^2\phi_k$ and $I_s^\beta\propto \sin^2\phi_k$ (Fig.~\ref{fig_AgTe} a,b). In analogy, $p$-polarized light projects out the $p_x$ character of band $|\alpha\rangle$ and delivers $ I_p^\alpha\propto \sin^2\phi_k$. In band $|\beta\rangle$, however, $s$-and $d$-channel interference again yields a $k$-dependent final state intensity $I_p^\beta \propto| \cos\phi_k (\sqrt{2}\widetilde{g}+\widetilde{f})\cos\alpha+i k_{\|} \delta_{sp}(\sqrt{2}\widetilde{g}-2\widetilde{f})\sin\alpha|^2$, which breaks the twofold rotational symmetry of the $\cos^2\phi_k$ if photo emission $s$- and $d$-channels are out of phase. This produces the oppositely oriented half moon structures in Fig.~\ref{fig_AgTe} (c,e) \cite{UenzelmannPhD,Uenzelmann_unpublished}, whose angular intensity distributions are fitted to the model in panels (d,f), and provide the relevant cross-section ratios and Coulomb phase shifts annotated to the figure.
\\

\textit{Discussion.}---Finally, let us discuss why -- despite the preceding arguments -- the PWA has served so well in describing ARPES intensity distributions, in particular also in \textit{orbital tomography} \cite{Bradshaw2015}: The latter is typically applied to small organic molecules, whose main element carbon is light, Coulomb final state effects fade out quickly with increasing photon energy ($\eta\propto Z/k \rightarrow 0$, $\widetilde{f}/\widetilde{g}\sim \text{const}$) and at least the angular part of the PWA holds. The investigated orbital character is almost entirely C $2p$, spin-orbit effects and OAM can thus be neglected and proper light polarization further limits orbital interference to a minimum -- ideally even suppresses one of the two photo emission channels altogether. 

However, it is exactly these reasons that render the PWA problematic in more complex systems. Intermixing of any additional orbital character will introduce two additional Coulomb wavelets that participate in final state interference, and if hybridization to heavy elements, e.g., to a metallic substrate is involved, Coulomb scattering will become significant. Orbital details obtained from a Fourier reconstruction in orbital tomography might then be meaningless.

Albeit it is not obvious if and how a similar reconstruction based on Coulomb waves could be implemented without bias (the outgoing photoelectron \textit{exit} wave is deformed by a \textit{per se} unknown potential), a detailed quantitative confrontation of model and experiment beyond the conceptional discussion in this work might still be feasible. This, however, crucially relies on well constructed initial states, both what concerns the angular part (that can be routinely obtained by downfolding of a Kohn-Sham eigenbasis obtained from density functional theory \cite{Mostofi2008}), but \textit{in particular} what concerns the radial part, which depends on a (less obvious) realistic description of Coulomb potentials close to the nuclei. If such an approach will turn out predictive for complex single- or even many electron systems has to be seen. What we have already shown so far, however, is that this simple \textit{Huygens principle of ARPES} has the potential to deliver ballpark figures of final state interference effects that so far required one-step photoemission calculations at the \textit{ab initio} level \cite{Scholz2013,Dauth2016,Bentmann2021}, while it maintains the computational ease and the priceless intuition of the PWA. 
\\\\
\textit{Acknowledgements.}---I thank Henriette Maa\ss, Maximilian \"{U}nzelmann, Hendrik Bentmann and Friedel Reinert of EP7, W\"{u}rzburg, for raising the problem of final state interference in BiAg$_2$ and AgTe, and for sharing their experimental data. Further, I thank Philipp Eck, Jonas Erhardt, Hans Kirschner, Peter Puschnig, Ralph Claessen and Phil Woodruff for helpful discussions and valuable feedback on this work. Funding support came from the Deutsche Forschungsgemeinschaft (DFG, German Research Foundation) under Germany’s Excellence Strategy through the W\"{u}rzburg-Dresden Cluster of Excellence on Complexity and Topology in Quantum Matter ct.qmat (EXC 2147, Project ID 390858490) as well as through the Collaborative Research Center SFB 1170 ToCoTronics (Project ID 258499086).

\bibliographystyle{apsrev4-1}

%


\end{document}